%% file: eos.tex
\documentclass[%
 reprint,
 amsmath,amssymb,
 aps,
 prd
]{revtex4-2}

\usepackage{graphicx}
\usepackage{float}
\usepackage{physics}

\newcommand{\hmu}{\hat{\mu}}

\begin{document}

\title{Resummed lattice QCD equation of state at \\ 
finite baryon density: strangeness neutrality and beyond}

\author{Szabolcs Bors\'anyi}
\affiliation{Department of Physics, Wuppertal University, Gaussstr.  20, D-42119, Wuppertal, Germany}

\author{Zolt\'an Fodor}
\affiliation{Pennsylvania State University, Department of Physics, University Park, PA 16802, USA}
\affiliation{Department of Physics, Wuppertal University, Gaussstr.  20, D-42119, Wuppertal, Germany}
\affiliation{Inst.  for Theoretical Physics, ELTE E\"otv\"os Lor\' and University, P\'azm\'any P. s\'et\'any 1/A, H-1117 Budapest, Hungary}
\affiliation{J\"ulich Supercomputing Centre, Forschungszentrum J\"ulich, D-52425 J\"ulich, Germany}
\affiliation{Physics Department, UCSD, San Diego, CA 92093, USA}

\author{Jana N. Guenther}
\affiliation{Department of Physics, Wuppertal University, Gaussstr.  20, D-42119, Wuppertal, Germany}

\author{Ruben Kara}
\affiliation{Department of Physics, Wuppertal University, Gaussstr.  20, D-42119, Wuppertal, Germany}

\author{Paolo Parotto}
\affiliation{Pennsylvania State University, Department of Physics, University Park, PA 16802, USA}

\author{Attila P\'asztor}
\affiliation{Inst.  for Theoretical Physics, ELTE E\"otv\"os Lor\' and University, P\'azm\'any P. s\'et\'any 1/A, H-1117 Budapest, Hungary}

\author{Claudia Ratti}
\affiliation{Department of Physics, University of Houston, Houston, TX 77204, USA} 

\author{K\'alm\'an Szab\'o}
\affiliation{Department of Physics, Wuppertal University, Gaussstr.  20, D-42119, Wuppertal, Germany}
\affiliation{J\"ulich Supercomputing Centre, Forschungszentrum J\"ulich, D-52425 J\"ulich, Germany}

\date{\today}

\begin{abstract}
We calculate a resummed equation of state with lattice QCD simulations at imaginary chemical potentials.
This work presents a generalization of the scheme introduced in Ref.~\cite{Borsanyi:2021sxv} to 
the case of non-zero $\mu_S$, focusing on the line of strangeness neutrality. 
We present results up to $\mu_B/T \leq 3.5$ on the strangeness neutral line $\left\langle S \right\rangle = 0$ in the temperature range $130\textrm{\ MeV} \leq T \leq 280$~MeV. We also extrapolate the finite baryon density equation of state to small non-zero 
values of the strangeness-to-baryon ratio $R=\left\langle S \right\rangle / \left\langle B \right\rangle$. 
We perform a continuum extrapolation using lattice simulations of the 4stout-improved staggered action 
with 8, 10, 12 and 16 timeslices.
\end{abstract}

\maketitle

\section{Introduction}
The properties of hot and dense strongly interacting matter 
are the subject of many theoretical and experimental studies. 
At zero baryon density, first principle lattice QCD simulations showed that 
the transition between deconfined and confined matter 
is a crossover~\cite{Aoki:2006we}. It is conjectured
that the crossover line in the 
temperature($T$)-baryochemical potential($\mu_B$) 
plane eventually turns into a line of first order transition
at a critical endpoint~\cite{Fukushima:2013rx,Kovacs:2016juc,Critelli:2017oub,Isserstedt:2019pgx,Bernhardt:2021iql,Gao:2020fbl}.

At finite baryon density, 
lattice simulations face the infamous complex action or sign problem~\cite{Aarts:2015tyj, Nagata:2021bru}.
Most results on finite density QCD therefore come from indirect methods, such as 
reweighting from zero chemical potential~\cite{Hasenfratz:1991ax,Fodor:2001au,
Fodor:2001pe,Fodor:2002km,Fodor:2004nz, Giordano:2020uvk,Giordano:2020roi}, Taylor expansion around zero chemical
potential
~\cite{Allton:2002zi,Allton:2005gk,Gavai:2008zr,Borsanyi:2011sw,
Borsanyi:2012cr,Bellwied:2015lba,Ding:2015fca,Bazavov:2017dus,Fodor:2018wul, 
Giordano:2019slo, Giordano:2019gev,Bazavov:2020bjn},
or analytic continuation from imaginary chemical potentials, where the 
sign problem is absent~\cite{deForcrand:2002hgr,DElia:2002tig,DElia:2009pdy, 
Cea:2014xva,Bonati:2014kpa,Cea:2015cya,Bonati:2015bha,
Bellwied:2015rza,DElia:2016jqh,Gunther:2016vcp,Alba:2017mqu,
Vovchenko:2017xad,Bonati:2018nut,Borsanyi:2018grb,
Borsanyi:2020fev,Pasztor:2020dur}.
More direct simulation approaches at finite density include certain 
reweighting techniques~\cite{Fodor:2007vv,Giordano:2020roi,Borsanyi:2021hbk}, 
the complex Langevin equation~\cite{Aarts:2009uq,Sexty:2013ica,Scherzer:2019lrh} 
and methods 
based on Lefschetz thimbles~\cite{Cristoforetti:2012su,Alexandru:2015xva,Fukuma:2019uot}. 
Results with these more direct methods for full QCD are for the 
moment nonexistent (for Lefschetz thimbles) or sparse (for complex Langevin and reweighting). 

Another approach to locate the critical endpoint is the analysis of 
data from heavy ion collision experiments.
An important tool for heavy ion phenomenology is the use of 
relativistic hydrodynamic simulations~\cite{Romatschke:2009im,Heinz:2013wva,Pratt:2015zsa}. These in turn 
need the equation of state as a theoretical 
input in the whole range of temperatures and densities 
covered by the experiment. 

The equation of state at zero density (or baryochemical potential $\mu_B$)
is known in the continuum limit from the crossover 
region~\cite{Borsanyi:2010cj, Borsanyi:2013bia, Bazavov:2014pvz} up to very high 
temperatures~\cite{Borsanyi:2016ksw} where it can be matched with results from 
resummed perturbation theory~\cite{Kajantie:2002wa,Andersen:2010wu,Andersen:2011sf}. 

The most straightforward way to extend the equation of state to 
finite chemical potentials is the use of a Taylor expansion in the 
chemical potential. For the pressure this reads:
\begin{equation}
    \frac{p}{T^4}\left( T, \hmu_B, \hmu_S, \hmu_Q \right) = \sum_{ijk} \frac{1}{i!j!k!} \chi^{BSQ}_{ijk}\left( T \right) \hat{\mu}_B^i \hat{\mu}_S^j \hat{\mu}_Q^k \rm{,}
\end{equation} 
where the generalized susceptibilities are defined as
\begin{equation}
    \label{eq:generalized_susceptibilities}
    \chi^{BSQ}_{ijk} = \frac{\partial^{i+j+k}\left( \hat{p} \right)}{\partial^i \hat{\mu}_B \partial^j \hat{\mu}_S \partial^k \hat{\mu}_Q}
\end{equation}
and the dimensionless baryochemical potential is $\hat{\mu}_B = \frac{\mu_B}{T}$, and similarly for the strangeness chemical potential $\mu_S$ and the electric charge chemical potential $\mu_Q$. The dimensionless pressure is $\hat{p} = \frac{p}{T^4}$.
The expansion coefficients $\chi^{BSQ}_{ijk}$ can be calculated either by direct simulations at zero chemical potential or by 
simulations at imaginary chemical potentials, followed by a fit. The Taylor coefficients are known up to fourth order 
in the continuum limit~\cite{Bellwied:2015lba, Ding:2015fca, Bazavov:2017dus}. At high temperatures, these coefficients match those calculated 
from resummed perturbation theory~\cite{Mogliacci:2013mca,Haque:2014rua,Haque:2020eyj}.
At finite lattice spacings, also the sixth and eights derivatives have been calculated, albeit with 
modest precision~\cite{Borsanyi:2018grb,Bazavov:2020bjn}.

Naive truncations of the Taylor expansion show undesirable 
properties above $\mu_B/T \gtrapprox (2-2.5)$: namely, unphysical oscillations in the 
equation of state. This is due to the sign structure of higher order coefficients starting from $\chi^{B}_6$, 
which is in turn caused by the $\hmu_B$ dependence of the crossover 
transition~\cite{Ratti:2007jf, Critelli:2017oub, Parotto:2018pwx,Borsanyi:2021sxv,Karthein:2021nxe}.
It appears that extrapolating through the crossover transition - as one is forced to 
do with Taylor expansions at a fixed temperature - is hard.

\begin{figure*}[t!]
    \begin{center}
        \includegraphics[width=0.950\linewidth]{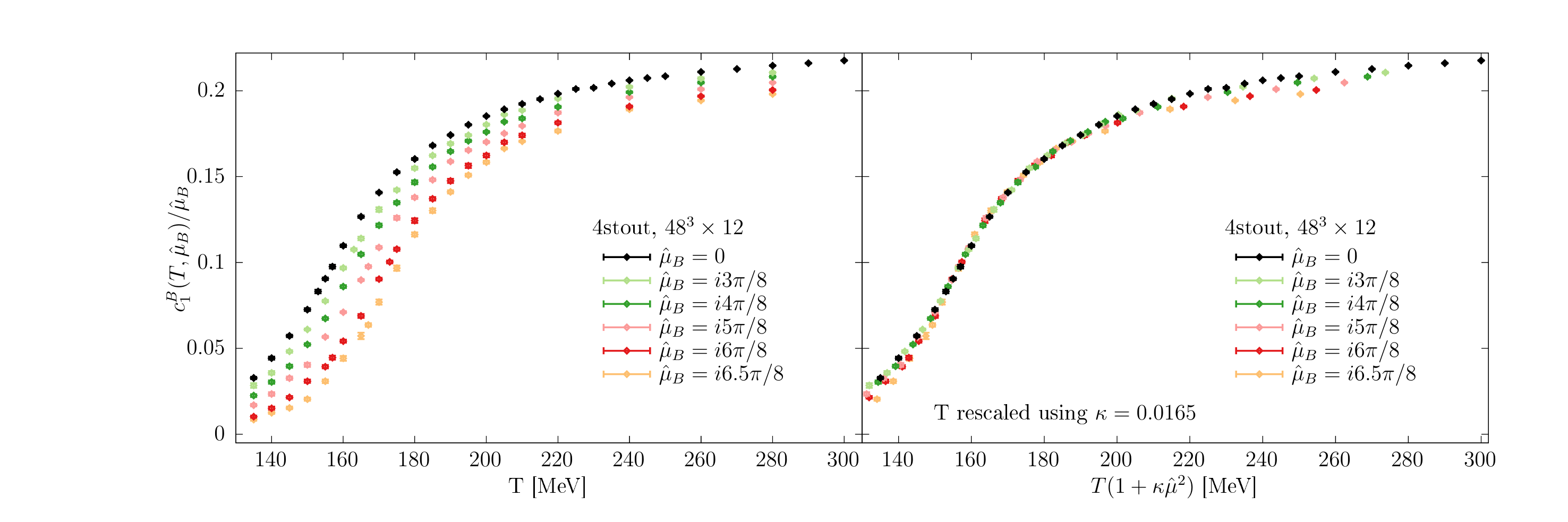}
    \end{center}
    \vspace{-0.6cm}
    \caption{
    \label{fig:collapse_oldscheme}        
Left: The total derivative $c^B_1$ on the 
strangeness neutral line from our imaginary chemical potential simulations. The data 
points at $\mu_B=0$ show the second derivative $\frac{\dd^2 \hat{p}}{\dd \hat{\mu}_B^2}$. 
Right: Same observables, with the temperature rescaled by a factor $1+\kappa\hat{\mu}_B^2$.
}
\end{figure*}

One can go beyond fixed order polynomials, by considering
resummations of the Taylor series. The classic example of this is 
Padé resummation, which has been used also for finite 
density QCD~\cite{Cea:2014xva, Datta:2016ukp, Pasztor:2020dur, Dimopoulos:2021vrk}.
While the convergence properties of Padé approximants are in general superior
to ordinary Taylor expansions, it is hard to tell which expansion will work
better at the low orders of the expansion available. 

A way to extend the chemical potential range accessible 
by indirect methods already at low orders of the approximation 
is the use of a physically motivated extrapolation scheme. 
Recently, 
in Ref.~\cite{Borsanyi:2021sxv} 
the authors of this paper introduced exactly 
such a scheme. It was motivated
by an important finding in recent 
studies of the crossover region at imaginary 
chemical potentials: the existence of an approximate scaling 
variable $T(1+\kappa \left( 1+ \kappa \hmu_B^2\right)$, 
where $T$ is the temperature, and $\mu_B$ 
the baryochemical potential. In this
variable, many observables - such as the chiral 
condensate or the baryon number-to-chemical potential 
ratio $\chi^B_1/\mu_B$ -  when measured at different 
fixed values of $\operatorname{Im} \mu_B/T$ as a 
function of temperature, collapse to the same 
curve~\cite{Borsanyi:2020fev,Borsanyi:2021sxv}.
Up to $\mu_B/T=1.5$ the existence of an approximate scaling variable 
has also been confirmed directly at a real baryochemical potential, 
by reweighting from the sign quenched 
ensemble~\cite{Borsanyi:2021hbk}. 

One might suspect that the existence of such an approximate 
scaling variable is ultimately related to criticality 
in the two-flavour chiral limit of QCD~\cite{HotQCD:2019xnw,Kotov:2021rah}. If the universal contribution to the equation 
of state is large, one expects a single scaling variable
- which is a combination of the quark mass, the temperature and the chemical potential - to govern most thermodynamic observables.
If in the low chemical potential region the strength of the 
transition is ultimately governed by the quark mass, one expects curves sensitive to criticality to not change shape 
significantly at small chemical potentials. This is also 
consistent with the observation that the width of the 
crossover transition is, to a good approximation, constant
at small $\mu_B$~\cite{Bonati:2015bha,Borsanyi:2020fev}.

Based on these observations, one can define 
the systematically improvable ansatz:
\begin{equation}
    \label{eq:oldscheme}
    F(T,\hat{\mu}_B) = F(T(1+\kappa^F_2(T) \hat{\mu}_B^2+\kappa^F_4(T) \hat{\mu}_B^4+\dots ) , 0)\rm{,}
\end{equation}
where $F$ is some observable of interest, of sigmoid shape in the temperature, 
such as $\chi^B_1/\hmu_B$.  The superscripts on the $\kappa^F_n$ denote that the 
expansion coefficients are different for different observables.
The approximate 
constant strength of the crossover transition at small chemical 
potentials is manifest in the approximate temperature independence of 
the coefficient $\kappa_2(T)$ in the crossover region.
A strengthening or weakening of the transition at larger chemical 
potentials could in turn manifest itself in a  non-trivial 
temperature dependence of the higher order expansion 
coefficients $\kappa_n$ in the crossover region. 

While in our previous study in Ref.~\cite{Borsanyi:2021sxv} we showed the 
QCD equation of state up to $\mu_B/T=3.5$ for the simplest case of 
$\mu_S=\mu_Q=0$, for the 
phenomenologically more relevant strangeness neutral 
line $\chi^S_1 = 0$, the equation of state from first
principle calculations is still only available as a fixed order Taylor expansion at fixed $T$, 
and it therefore covers 
a rather limited range in the chemical potentials,
due to the aforementioned unphysical oscillations.

Here, we extend our previous investigations using our novel resummation scheme, 
conducted at zero strangeness chemical potential $\mu_S=0$ in 
Ref.~\cite{Borsanyi:2021sxv} to the strangeness neutral line, and present 
results up to $\mu_B/T=3.5$ in this setting. This equation of state can be used in
those hydrodynamic simulations where local strangeness neutrality is enforced everywhere.

We also introduce a further component of the resummation scheme - the Stefan-Boltzmann 
correction discussed in the next section. This will 
improve the convergence of our ansatz at high temperatures - 
where the aforementioned approximate scaling does not hold. 

Finally, we go beyond strangeness neutrality by calculating - at finite real 
$\mu_B$ - the expansion coefficients needed to calculate the equation of state at 
small values of the strangeness-to-baryon number 
ratio:
\begin{equation}
    R \equiv \frac{\left\langle S \right\rangle}{\left\langle B \right\rangle} = \frac{\chi^S_1}{\chi^B_1} \rm{.}
\end{equation}
This allows one to relax the condition of local strangeness neutrality in hydrodynamic 
simulations. This is necessary, as in a heavy ion collision, only global strangeness
neutrality is guaranteed, and local charge fluctuations can be 
large.

The structure of the paper is as follows. In Sec. II we briefly 
describe the strangeness neutrality condition. In Sec. III. we
formally 
define our extrapolation ansatz on the strangeness neutral line,
including the Stefan-Boltzmann improvement. In Sec. IV
we describe our method to determine the extrapolation coefficients.
In Sec. V we use the coefficients determined in Sec. IV to 
calculate thermodynamics at finite real chemical potentials 
both on the strangeness neutrality line and its vicinity.

\section{Strangeness neutrality}
When enforcing strangeness neutrality, two sets of conditions are widely used in the literature. 
In the first, one sets $\mu_Q=0$ and defines a curve in the $\mu_B-\mu_S$ plane by the condition
$\chi^S_1=0$. By differentiation, this implies:
\begin{equation}
    \frac{\dd \mu_S}{\dd \mu_B} = -\frac{\chi^{BS}_{11}}{\chi^S_2} \rm{.}
\end{equation}
On this line, total derivatives with respect to the baryochemical potential read
\begin{equation}
    \frac{\dd}{\dd\hat{\mu}_B} 
    = \frac{\partial}{\partial \hat{\mu}_B} + \frac{\dd \hat{\mu}_S}{\dd \hat{\mu}_B} \frac{\partial}{\partial \hat{\mu}_S} 
    = \frac{\partial}{\partial \hat{\mu}_B} - \frac{\chi^{BS}_{11}}{\chi^S_2}     \frac{\partial}{\partial \hat{\mu}_S} \rm{.}
\end{equation}
We denote the total derivatives of the dimensionless pressure with respect to the baryochemical potential along 
the line $\mu_Q=0$ and $\chi^S_1=0$ as:
\begin{equation}
        \begin{aligned}
        c^B_n(T,\hat{\mu}_B) \equiv \frac{\dd^n \hat{p}(T,\hat{\mu}_B) }{\dd \hat{\mu}_B^n} \Bigg|_{\substack{\mu_Q=0 \\ \chi^S_1=0} }\rm{.}
        \end{aligned}
\end{equation}
In this scheme, the leading coefficient
\begin{equation}
c^B_1(T,\hat{\mu}_B) = \chi^B_1 - \frac{\chi^{BS}_{11}}{\chi^S_2}\chi^S_1 = \chi^B_1
\end{equation}
gives the net baryon density.

Since a vanishing electric charge chemical potential is equivalent to zero isospin, this condition 
is equivalent to $\chi^Q_1 = 0.5 \chi^B_1$. While this is a much better approximation for the 
conditions of a heavy ion collision than simply taking $\mu_S=\mu_Q=0$, there is a further correction 
which can be taken into account: the slight isospin imbalance of the colliding nuclei (typically lead or gold).
This amounts to the two constraints $\chi^S_1=0$ and $0.4\chi^B_1=\chi^Q_1$, which define a curve in the 
$\mu_B-\mu_S-\mu_Q$ space. Along this curve the 
total derivatives are: 
\begin{equation}
    \frac{\dd}{\dd\hat{\mu}_B} 
    = \frac{\partial}{\partial \hat{\mu}_B} + \frac{\dd \hat{\mu}_S}{\dd\hat{\mu}_B} \frac{\partial}{\partial \hat{\mu}_S} 
    + \frac{\dd \hat{\mu}_Q}{\dd \hat{\mu}_B} \frac{\partial}{\partial \hat{\mu}_Q} \rm{,} 
\end{equation}
where $\frac{\dd \hat{\mu}_S}{\dd \hat{\mu}_B}$ and $\frac{\dd \hat{\mu}_Q}{\dd \hat{\mu}_B}$ are given by the solution of the 
constraints: 
\begin{equation}
    \begin{aligned}
        \chi^{BS}_{11} &+ \chi^{SQ}_{11} \frac{\dd \hat{\mu}_Q}{\dd \hat{\mu}_B} + \chi^{S}_{2} \frac{\dd \hat{\mu}_S}{\dd \hat{\mu}_B} = 0 \\
        \chi^{BQ}_{11} &+ \chi^{SQ}_{11} \frac{\dd \hat{\mu}_S}{\dd \hat{\mu}_B} + \chi^{Q}_{2} \frac{\dd \hat{\mu}_Q}{\dd \hat{\mu}_B} = \\ 
        & 0.4 \left(  \chi^B_2 + \chi^{BS}_{11} \frac{\dd \hat{\mu}_S}{\dd \hat{\mu}_B} + \chi^{BQ}_{11} \frac{\dd \hat{\mu}_Q}{\dd \hat{\mu}_B} \right)
    \end{aligned}
\end{equation}
Along this line, total derivatives will be denoted:
\begin{equation}
    \begin{aligned}
    d^B_n(T,\hat{\mu}_B) \equiv \frac{\dd^n \hat{p} (T,\hat{\mu}_B)}{\dd \hat{\mu}_B^n} \Bigg|_{\substack{\chi^Q_1 = 0.4 \chi^B_1 \\ \chi^S_1=0}} \rm{.}
    \end{aligned}
\end{equation}

For simplicity, through most of this manuscript, we will use the first set of conditions 
with $\mu_Q=0$. In Sec. IV.2., we will consider the difference between the two schemes 
in the leading order of the Taylor expansion - i.e. we will calculate $c^B_2(T,0)$
and $d^B_2(T,0)$ and their temperature derivatives, at which point we will also discuss 
the rationale for our choice of the first setting with $\mu_Q=0$.

\begin{figure*}[t!]
    \begin{center}
        \includegraphics[width=0.950\linewidth]{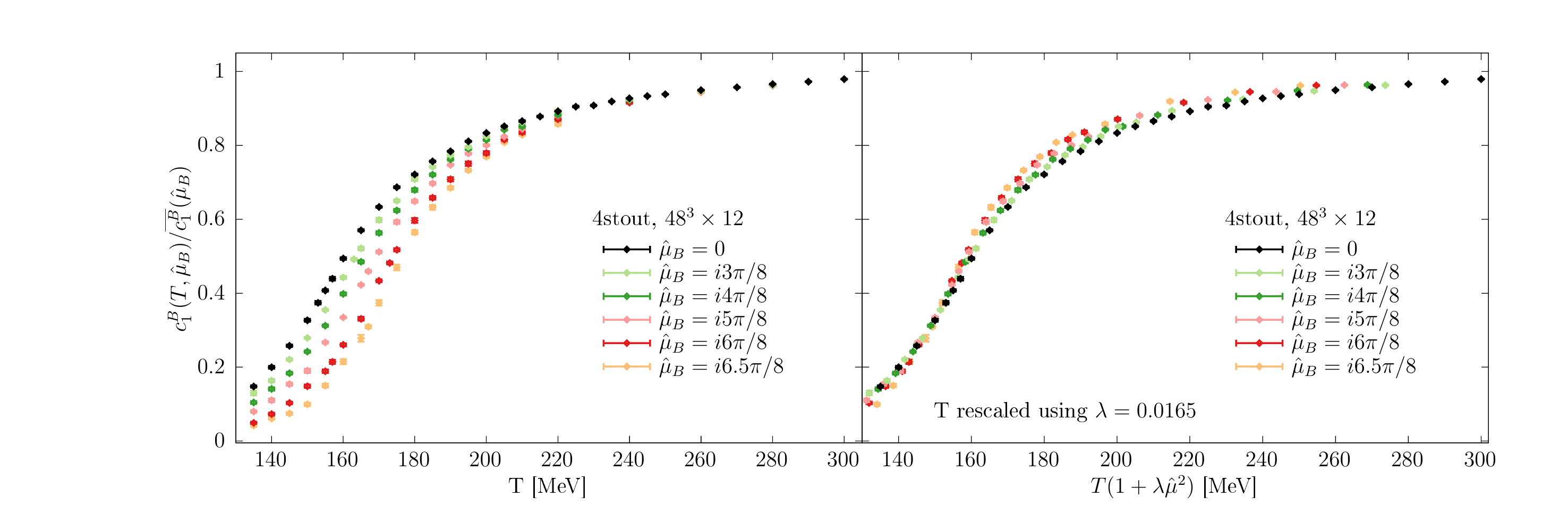}
    \end{center}
    \vspace{-0.6cm}
    \caption{
    \label{fig:collapse_SBscheme}        
Left: The scaled total derivative $c^B_1/\hmu_B$ on the 
strangeness neutral line from our imaginary chemical potential simulations, divided by its 
chemical potential dependent Stefan-Boltzmann limit. The data points at $\mu_B=0$ show the 
second derivative $c^B_2$ divided by its Stefan-Boltzmann limit. 
Right: Same observables, with the temperature rescaled by a factor $1+\lambda\hat{\mu}_B^2$.
}
\end{figure*}

\section{The extrapolation scheme
\label{sec:scheme}
}

Before describing our improved extrapolation ansatz, we note that 
the ansatz given by Eq.~\eqref{eq:oldscheme}, introduced 
in~\cite{Borsanyi:2021sxv} for $\mu_S=0$, would also work at strangeness neutrality.
The existence of the approximate scaling variable on the 
strangeness neutral line is shown in Fig.~\ref{fig:collapse_oldscheme}
for the quantity $c^B_1 / \mu_B$, where on the left panel we show the data points of our simulations 
for a $48^3 \times 12$ lattice, while on the right panel we show the same data points as a function 
of a rescaled temperature $T(1 + \kappa \hmu^2)$.
Notice that the collapse plot with a constant 
$\kappa$ does not work quite as well
at high temperatures. Indeed, one does not expect an approximate scaling variable outside the crossover range. 
Our scheme can still incorporate this behavior by the temperature dependence of 
the $\kappa_n$ coefficients. In fact, with the ansatz given by 
Eq.~\eqref{eq:oldscheme} the coefficient $\kappa_2$ grows at 
high temperatures (see later).

One shortcoming of this scheme 
is that the region of applicability 
is restricted by the Stefan-Boltzmann limit of the right 
hand side of Eq.~\eqref{eq:oldscheme}. When the 
quantity $F(T,\hat{\mu}_B)$ becomes larger than its infinite temperature limit at $\mu_B=0$, the ansatz in 
Eq.~\eqref{eq:oldscheme} must break down. It is easy to address this shortcoming however, using the scheme only for observables $F$
that have an infinite temperature limit that is independent of $\hat{\mu}_B$. Given an observable that does not possess this property, one 
can easily construct another observable, by simply dividing by its own Stefan-Boltzmann limit:
\begin{equation}
F(T,\hat{\mu}_B) \to \frac{F(T,\hat{\mu}_B)}{\overline{F}(\hat{\mu}_B)} \rm{,}
\end{equation}
where the Stefan-Boltzmann limits~\footnote{Throughout this work, we always use the continuum Stefan-Boltzmann limits.} are denoted by
\begin{equation}
    \label{eq:corrected_observable}
    \begin{aligned}
        \overline{F}(\hat{\mu}_B) &= \lim_{T \to \infty} F(T,\hat{\mu}_B) 
        \rm{.} 
    \end{aligned}
\end{equation}
By construction, the ratio on the right hand side of 
Eq.~\eqref{eq:corrected_observable} has an infinite 
temperature limit equal to one, at all values of $\hmu_B$. By using the ansatz from 
Eq.~\eqref{eq:oldscheme} on this Stefan-Boltzmann corrected 
observable, we arrive at our new scheme, given by
\begin{equation}
    \label{eq:ansatz_master}
    \frac{F(T,\hat{\mu}_B)}{\overline{F}(\hat{\mu}_B)} = \frac{F(T_F',0)}{\overline{F}(0)} \rm{,}
\end{equation}
where the temperature on the right hand side is expanded as 
\begin{equation}
    T_F' = T \left( 1 + \lambda^F_2(T) \hat{\mu}_B^2  + \lambda^F_4(T) \hat{\mu}_B^4 + \dots \right) \rm{.}
\end{equation}
As is shown in Fig.~\ref{fig:collapse_SBscheme}, this Stefan-Boltzmann correction does not spoil the collapse 
plot in the approximate scaling variable, meaning that the fast convergence of the scheme in the crossover 
region is maintained, with a $\lambda_2(T)$ coefficient that is approximately constant in the crossover range. 
The limitation at high temperature is however removed. Furthermore, as
can be seen on the left 
panel of Fig.~\ref{fig:collapse_SBscheme},
the coefficients $\lambda_n$ must go to zero at high temperatures, as the data points for 
the different imaginary chemical potentials almost overlap.
This is in contrast to the scheme of Eq.~\eqref{eq:oldscheme}, where $\kappa_2$ grows at high temperatures.

For any finite order in the expansion in the $\lambda_n$, 
Eq.~\eqref{eq:ansatz_master} generates an infinite 
number of terms in the Taylor expansion of the 
quantity $F$, thus the ansatz achieves a particular resummation 
of the Taylor expansion. As discussed in the introduction, this 
resummation is expected to converge fast in the crossover region,
where the strength of the transition stays approximately constant. 

In this work, we will consider three different observables $F$. First, we consider the normalized net baryon density
$F = c^B_1/\hmu_B$. 
By noticing that $\lim_{\hat{\mu}_B \to 0} \frac{c^B_1(T,\hmu_B)}{\hat{\mu}_B} = c^B_2(T,0)$,
Eq.~\eqref{eq:ansatz_master} becomes:
\begin{equation}
    \label{eq:ansatz_BB}
    \frac{c^B_1(T,\hat{\mu}_B)}{\overline{c^B_1}(\hat{\mu}_B)} = 
    \frac{c^B_2(T_{BB}',0)}{\overline{c^B_2}(0)}\rm{,}
\end{equation}
where the infinite temperature limits of $c^B_1$ and $c^B_2$ are denoted $\overline{c^B_1}$ and $\overline{c^B_2}$ respectively, and
\begin{equation}
    \label{eq:TprimeBB}
    T_{BB}' \equiv T(1+\lambda_2^{BB}(T) \hat{\mu}_B^2+\lambda_4^{BB}(T) \hat{\mu}_B^4+\dots) \rm{.}
\end{equation}
The Stefan-Boltzmann limits are easily obtained:
\begin{eqnarray}
\overline{c^B_1}(\hmu_B)&=&
\hmu_B\overline{c^B_2}(0)+\hmu_B^3\overline{c^B_4}(0)\nonumber\\
\overline{c^B_2}(0)&=&\frac{2}{9}\qquad
\overline{c^B_4}(0)=\frac{4}{27\pi^2}
\end{eqnarray}
The $c_2^B(T,0)$ function at $\mu_B=0$ together with the temperature-dependent
coefficients in Eq.~(\ref{eq:TprimeBB}) are sufficient to extrapolate
the strangeness-neutral equation of state to finite baryon density as
we will show in Section.~\ref{sec:thermo}.

Second, we will consider the normalized strangeness chemical potential that
is needed to realize the $\chi^S_1\equiv0$ condition in a grand canonical ensemble: $F=\rm{M}(T,\hat{\mu}_B)\equiv\frac{\hat{\mu}_S}{\hat{\mu}_B}\left( T, \hat{\mu}_B\right)$.
Since
\begin{equation}
    \lim_{\hat{\mu}_B \to 0} \rm{M}(T,\hat{\mu}_B) = -\frac{\chi^{BS}_{11}(T,0)}{\chi^S_2(T,0)} \equiv \rm{M}(T,0) \rm{,}
\end{equation}
Eq.~\eqref{eq:ansatz_master} becomes:
\begin{equation}
    \label{eq:ansatz_BS}
    \frac{\rm{M}(T,\hat{\mu}_B)}{\overline{\rm{M}}(\hat{\mu}_B)} = 
    \frac{\rm{M}(T_{BS}',0)}{\overline{\rm{M}}(0)}\rm{,}
\end{equation}
with the Stefan Boltzmann limit $\overline{\rm{M}}(\hat{\mu}_B) = \lim_{T \to \infty}  \rm{M}(T,\hat{\mu}_B)$
 and
\begin{equation}
    T_{BS}'=T(1+\lambda_2^{BS}(T) \hat{\mu}_B^2+\lambda_4^{BS}(T) \hat{\mu}_B^4+\dots)\rm{.}
\end{equation}

\begin{figure}[h!]
    \begin{center}
        \includegraphics[width=0.98\linewidth]{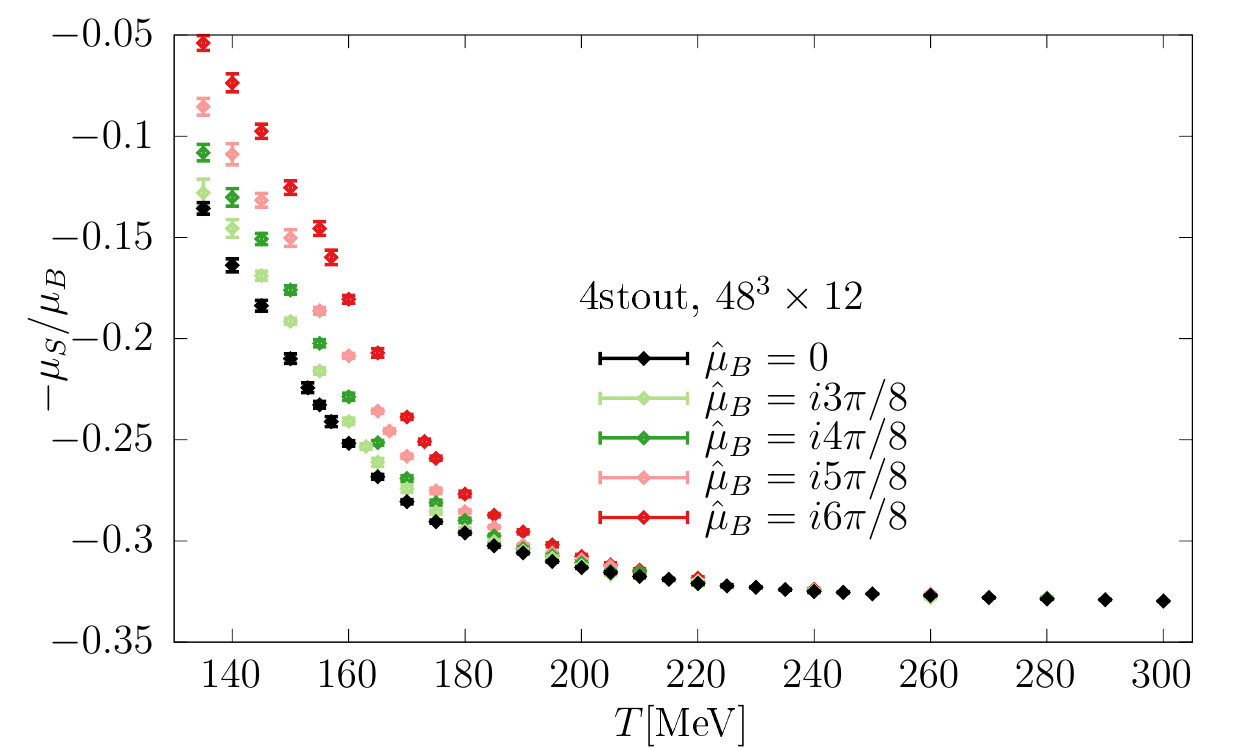}  
        \includegraphics[width=0.98\linewidth]{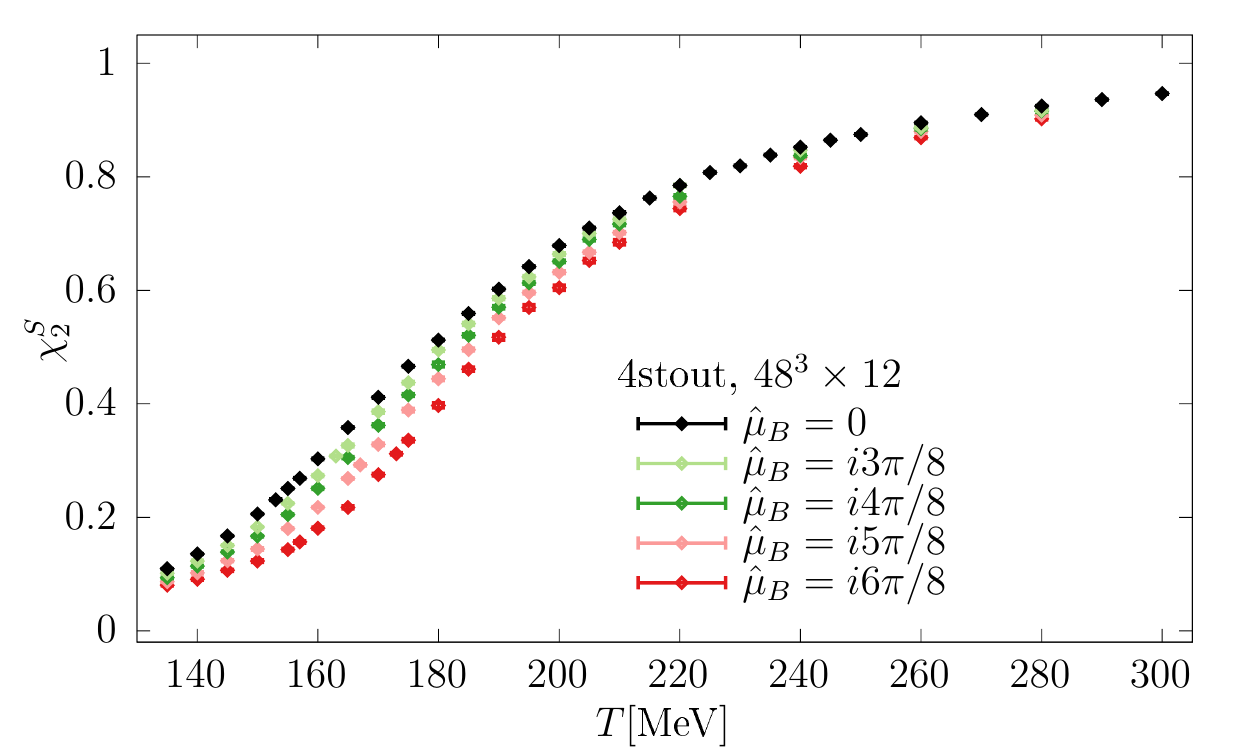}
    \end{center}
    \vspace{-0.3cm}
    \caption{
    \label{fig:data_BS_SS}
        The strangeness to baryon chemical potential ratio (upper panel) and the
        strangeness susceptibility (lower panel) at simulated imaginary baryochemical potentials 
        on our $48^3 \times 12$ ensembles. 
    }
\end{figure}

Finally, we will also consider $F=\chi^S_2$, and denote its Stefan-Boltzmann limit by $\overline{\chi^S_2}$. For this observable,
Eq.~\eqref{eq:ansatz_master} reads:
\begin{equation}
    \label{eq:ansatz_SS}
    \frac{\chi^S_2(T,\hat{\mu}_B)}{\overline{\chi^S_2}(\hat{\mu}_B)} = 
    \frac{\chi^S_2(T_{SS}',0)}{\overline{\chi^S_2}(0)}\rm{,}
\end{equation}
where 
\begin{equation}
    T_{SS}'=T(1+\lambda_2^{SS}(T) \hat{\mu}_B^2+\lambda_4^{SS}(T) \hat{\mu}_B^4+\dots)\rm{.}
\end{equation}
Note that, at strangeness neutrality, the Stefan-Boltzmann limits of
$\rm{M}$ and $\chi^S_2$ are independent of $\hat{\mu}_B$,
\begin{equation}
\overline{\rm{M}}(\hat{\mu}_B)=\hat{\mu}_B \overline{\rm{M}}(0)\,,\qquad
\overline{\chi^S_2}(\hat{\mu}_B)=1\,,
\end{equation}
 thus 
$\kappa_n^{BS}=\lambda_n^{BS}$ and $\kappa_n^{SS}=\lambda_n^{SS}$.
This concludes the definition of our scheme for all observables of interest.
We show the lattice data for $\rm{M}$ and $\chi^S_2$ on our $48^3 \times 12$ ensembles 
in Figure~\ref{fig:data_BS_SS}.

\begin{figure}[h!]
    \begin{center}
        \includegraphics[width=0.98\linewidth]{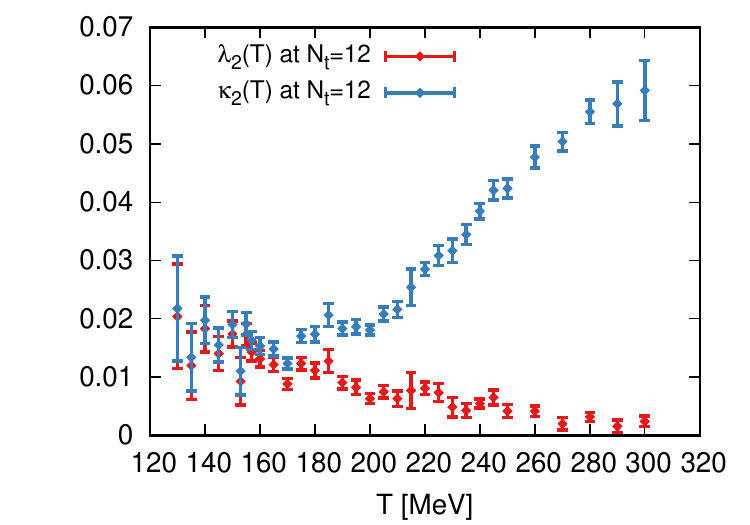}
    \end{center}
    \vspace{-0.7cm}
    \caption{
        \label{fig:SB_ornot_SB}
        The $\kappa_2 \equiv \kappa^{BB}_{2}$ coefficients obtained in the 
        scheme of Eq.~\eqref{eq:oldscheme} without the Stefan-Boltzmann correction (in red), 
        compared to the coefficients $\lambda_2 \equiv \lambda^{BB}_2$ obtained in the 
        scheme of Eq.~\eqref{eq:ansatz_master} with the Stefan-Boltzmann correction (in black) 
        on our $48^3 \times 12$ lattices.
    }
\end{figure}

Finally, in order to illustrate the advantageous properties of the Stefan-Boltzmann correction mentioned above, we compare 
the $\kappa^{BB}_2$
and $\lambda^{BB}_2$ coefficients obtained without and with the Stefan-Boltzmann correction respectively, in Fig. ~\ref{fig:SB_ornot_SB} for our $48^3 \times 12$ lattices.

\section{Determination of the expansion coefficients}

\subsection{Lattice setup}
We perform simulations with $N_f=2+1+1$ dynamical quark flavours, with physical light, strange and charm quark 
masses. We use staggered fermions with fat links constructed with 4 steps of stout 
smearing~\cite{Morningstar:2003gk} with the smearing parameter $\rho=0.125$ 
and a tree-level Symanzik-improved gauge action. This discretization of the 
QCD action was first used in Ref.~\cite{Bellwied:2015lba}, where information about the
line of constant physics can be found. For the scale setting we use either the pion
decay constant $f_\pi= 130.41$MeV~\cite{Tanabashi:2018oca}, or the Wilson flow based 
$w_0=0.1725$fm scale introduced in Ref.~\cite{Borsanyi:2012zs}. Taking into account the difference 
between the two scale settings is part of our systematic error analysis. We use lattices 
of temporal extent $N_\tau=8, 10, 12$ ad $16$ to perform a continuum limit. The spatial volume is 
given by the aspect ratio of $LT=4$. We performed simulations for 
imaginary baryochemical potentials given 
by $\operatorname{Im} \hat{\mu}_B \frac{8}{\pi}= 0,3,4,5,6,6.5$. In addition, for the $N_\tau=12$ lattices 
we also have data at $\operatorname{Im} \hat{\mu}_B \frac{8}{\pi}=5.5$. 
Strangeness neutrality was enforced on our imaginary chemical potential ensembles via the procedure discussed in Ref.~\cite{Borsanyi:2020fev}.

\subsection{$c^B_2$, $d^B_2$ and their $T$ derivative at $\mu_B=0$}

In order to determine thermodynamics at finite real chemical potentials, the right hand side of 
Eq.~\eqref{eq:ansatz_BB}, i.e. $c^2_B(T,0)$ must be known. For some quantities, like the 
entropy, its temperature derivative is also needed. We describe the determination of this quantity 
and its derivative here.

Given the second order quark number susceptibilities at vanishing chemical potentials, one can 
express $c^B_2(T,0)$ as 
\begin{equation}
    c^{B}_{2}(T,0) = \chi^B_2 - \frac{\left( \chi^{BS}_{11} \right)^2}{\chi^S_2} \rm{,}
\end{equation}
which we continuum extrapolate using our $N_\tau=10,12$ and $16$ lattices (for $c^B_2(T)$ and $\rm{M}(T)$) or  $N_\tau=12,16$ and $20$ (for $\chi^S_2(T)$),
using a tree level improvement when applicable~\cite{Bellwied:2015lba}.
For the calculation of the 
temperature derivative we use the same procedure as in Ref.~\cite{Borsanyi:2021sxv}. A procedure which we 
also describe here, to make the manuscript self contained.

The derivatives are obtained by fitting the data with suitable ansätze, which address then differentiated.
We first divide the temperature range into two parts:
the crossover region and the high temperature region. For the low temperature region we interpolate 
$c^B_2$ with basis splines with nodes points in the range $T \in \left[ 130\rm{MeV},300\rm{MeV}\right]$. 
The splines are cubic. We perform a joint fit of the splines in temperature and in $1/N_\tau^2$.
The ansatz reads:
\begin{equation}
    c^2_{B}(T,0) = \sum_{i=1}^n \alpha_i b_i(T) + \frac{1}{N_\tau^2} \sum_{i=1}^{n} \beta_i b_i(T).
\end{equation}
Here, the basis splines satisfy $b_i(T_j)=\delta_{ij}$ where the $T_j$ with $1 \leq j \leq n$ are the node points for the given basis function. To estimate the systematics of the interpolation, we combine results from four sets of the 
node points $T_j$. The systematic error estimation also includes the difference between the $f_\pi$ and $w_0$ bases scale setting. 
We only kept fits with a $Q$ value higher than 5\%, these then we combined
with uniform weights.  The $c^B_2(T)$ and $\rm{M}(T)$ fits were constrained at
temperatures below $130$MeV to match predictions from the hadron resonance
gas (HRG) model. 

\begin{figure}[t!]
    \begin{center}
        \includegraphics[width=0.95\linewidth]{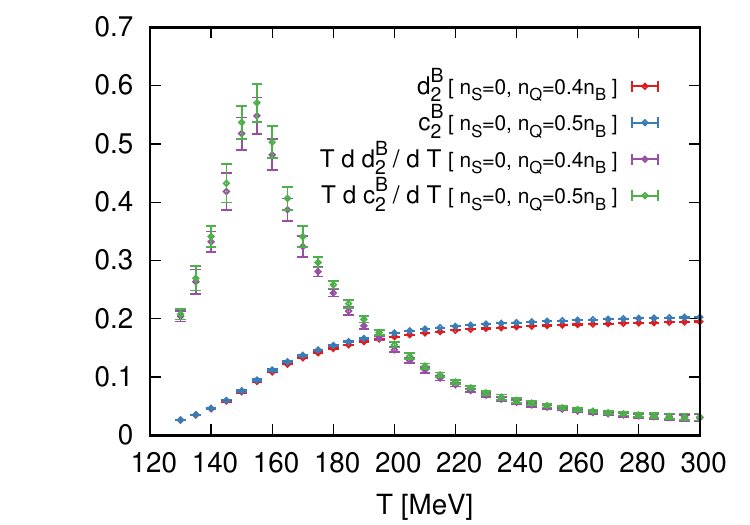}
    \end{center}
    \vspace{-0.5cm}
    \caption{
    \label{fig:c2B_plusderiv}        
    $c^B_2(T,\hat{\mu}_B=0)$, $d^B_2(T,\hat{\mu}_B=0)$ and their logarithmic temperature derivatives
    in the continuum limit, as extrapolated from our $N_\tau=10,12$ and $16$ lattices.
    }
\end{figure}

In the large temperature range we performed a polynomial fit in $1/T$. The known Stefan-Boltzmann limit is 
not enforced in the constant term of the polynomial. Hence, 
the fitted value of the constant is not equal 
to the known infinite temperature limit, and  our fit only 
allows for interpolation in the 
range where we have lattice data. The region where the 
two ansätze overlap give consistent results in the temperature range between $200$MeV and $280$MeV for both $c^B_2$ and its $T$ derivative. In the final result, 
we simply concatenate the two results at $T=250$MeV. Final results for $c^B_2(T,0)$ and its logarithmic temperature 
derivative are shown in Fig.~\ref{fig:c2B_plusderiv}.

We also performed the same analysis for $d^B_2(T,0)$, corresponding to $\chi_Q = 0.4 \chi_B$. The results for this 
quantity and its temperature derivative are also shown in Fig.~\ref{fig:c2B_plusderiv}. At large temperatures, 
there is a small but statistically significant difference between $d^B_2$ and $c^B_2$. The difference of these 
Taylor coefficients leads to a small difference between the leading order chemical potential dependence in these two
cases for high temperatures. The next corrections, corresponding to the $\lambda^{BB}_n$ coefficients of our 
resummation scheme, would probably also slightly differ in the two cases, but our lattice results are 
not yet precise enough to detect this difference. Therefore, we go on with the $\mu_Q=0$ setting for simplicity.

\begin{figure}[t!]
    \begin{center}
        \includegraphics[width=0.95\linewidth]{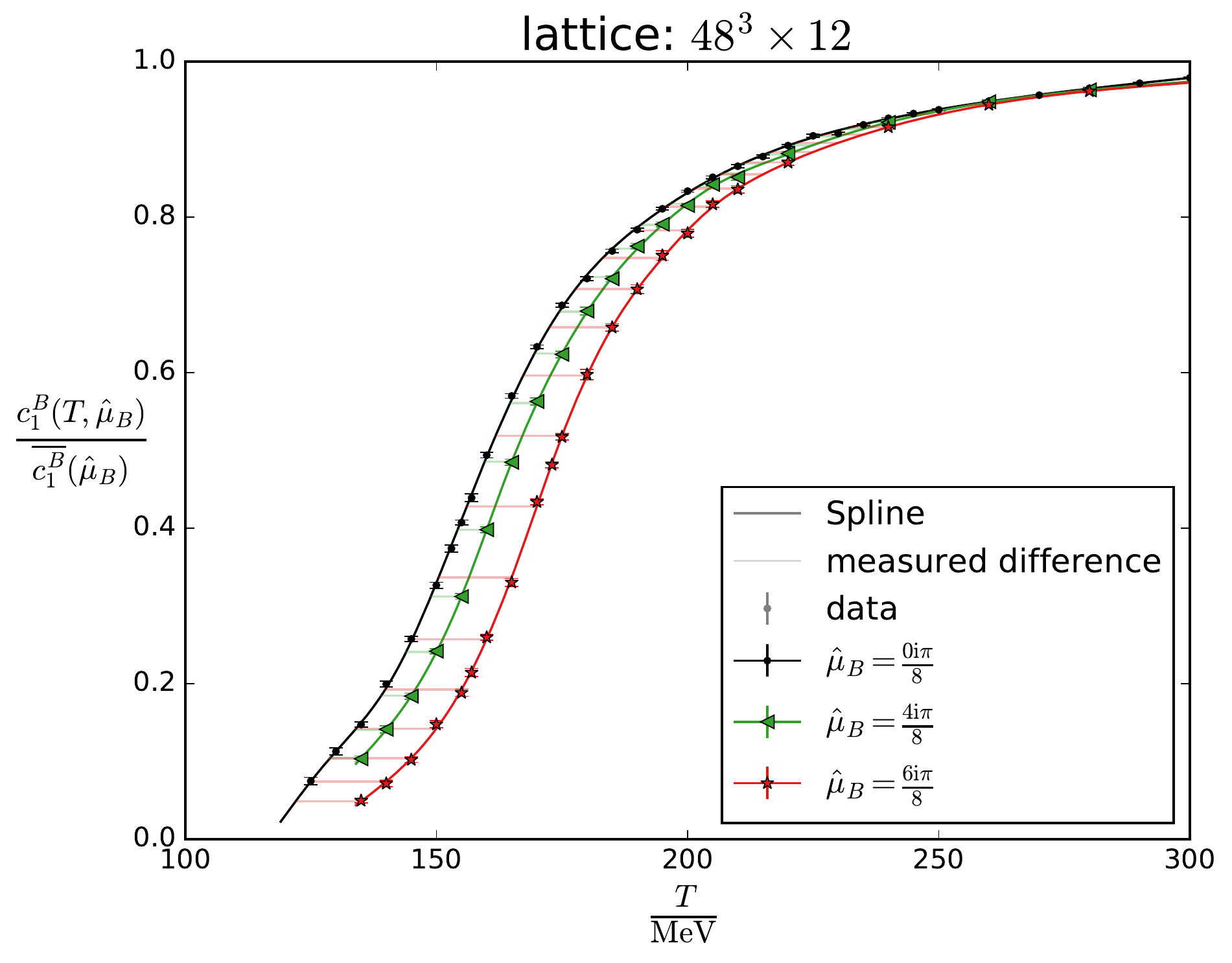}
    \end{center}
    \vspace{-0.5cm}
    \caption{
    \label{fig:input_Tprime}   
    The determination of the rescaled temperatures $T_{BB}'$ by spline fits to the data at zero and imaginary 
    baryochemical potential. Solid lines represent the spline fits to the lattice data, while transparent lines 
    represent the shifts $\Delta T_{BB} = T'_{BB}-T$extracted from the splines.
    }
\end{figure}

\begin{figure}[t!]
    \vspace*{-1cm}
    \begin{center}
        \includegraphics[width=0.95\linewidth]{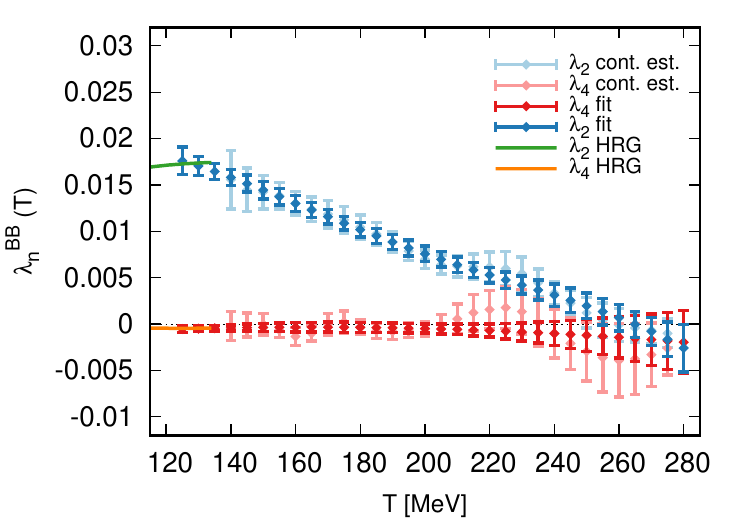}
        \includegraphics[width=0.95\linewidth]{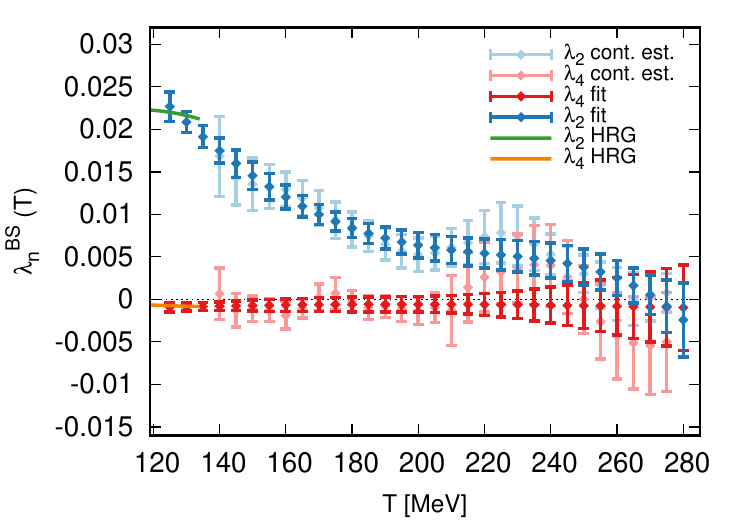}
        \includegraphics[width=0.95\linewidth]{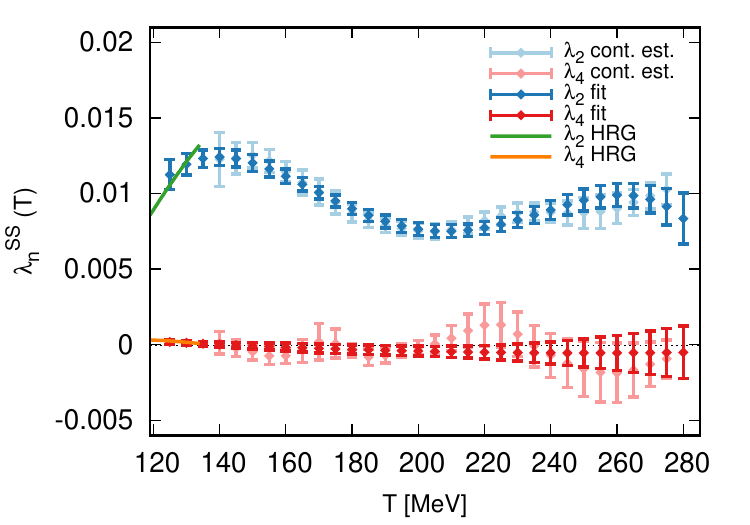}
    \end{center}
    \vspace{-8mm}
    \caption{
    \label{fig:lambdan}   
    Expansion coefficients $\lambda^{BB}_{n}$ (top panel), $\lambda^{BS}_{n}$ (middle panel) 
    and $\lambda^{SS}_{n}$ (bottom panel) in the continuum. Also shown are the fits used to 
    estimate the temperature derivative of these coefficients, as well as predictions  
    of the hadron resonance gas model, which we use to constrain the fits at low 
    temperatures.
    }
\end{figure}

\subsection{Analysis of the coefficients $\lambda_2$ and $\lambda_4$}

The analysis proceeds in the same way for all three observables we study.  
Denoting by $A$ either one of the observables $c^B_1$, $\rm{M}=\frac{\mu_S}{\mu_B}$ and $\chi^S_2$ and calling 
$B$ one of $c^B_2$, $-\frac{\chi^{BS}_{11}}{\chi^S_2}$ or $\chi^S_2$ respectively,
and denoting the corresponding Stefan-Boltzmann corrected observables by 
$\tilde{A} = A / \overline{A}$ and $\tilde{B} = B / \overline{B}$ respectively, 
our extrapolation ansatz is defined as
\begin{equation}
    \label{eq:solvethis}
    \tilde{A}(T,\hat{\mu}_B) = \tilde{B}(T', 0) \rm{.}
\end{equation} 
In the first step of the analysis, a spline interpolation is performed for $\tilde{A}$ at finite 
imaginary $\mu_B$ and for $\tilde{B}$ at $\mu_B=0$. 
In the second step, these splines are used for each imaginary $\hat{\mu}_B$, and several values 
of $\tilde{A}(T,\hat{\mu}_B)$ to determine a $T'$ that solves Eq.~\eqref{eq:solvethis}.
This procedure is illustrated in Fig.~\ref{fig:input_Tprime}. After finding the values of $T'$,
we construct the quantity
\begin{equation}
    \Pi(T,\hat{\mu}_B,N_\tau) = \frac{T'(T,\hat{\mu}_B,N)-T}{T \hat{\mu}_B} \rm{.}
\end{equation}
As $\lim_{\hat{\mu}_B \to 0} \Pi = \lambda_2^{ij}$, with $ij$ being either one of $BB$, $BS$ or $SS$, 
we can also add a data point at $\mu_B=0$ by utilizing the formulas connecting the ordinary Taylor 
coefficients defined in Eq.~\eqref{eq:generalized_susceptibilities} to the $\lambda_n$ coefficients
defined in Sec.~\ref{sec:scheme}. For reference, these formulas are listed 
in the Appendix.
In the third step, we fit the quantity $\Pi(T,\hat{\mu}_B,N_\tau) $ with an ansatz that is
linear in $1/N_\tau^2$ and either linear or parabolic in $\mu_B^2$ - i.e.:
\begin{equation}
    \begin{aligned}
    \Pi(T,\hat{\mu}_B,N_\tau) &= \lambda^{A}_2 + \lambda^{A}_4 \hat{\mu}_B^2 + \lambda^{A}_6 \hat{\mu}_B^4 \\ 
                              &+\frac{1}{N_\tau^2} \left( \alpha^A + \beta^A \hat{\mu}_B^2 + \gamma^A \hat{\mu}_B^4 \right)
    \end{aligned}
\end{equation}
where we either fix $\lambda^{A}_6 = \gamma^A = 0$ or leave both as free parameters in the fit.

Systematic errors were estimated by combining several fits with uniform weights, as 
long as their $Q$ value is above $1\%$. 
The different choices in the analysis procedure include:
\begin{itemize}
    \item 3 different sets of spline node points at $\mu_B$=0
    \item 2 different sets of spline node points at finite imaginary $\mu_B$
    \item $w_0$ or $f_\pi$ based scale setting
    \item 2 different chemical potential ranges in the global fit: $\hmu_B \leq 5.5$ or $\hmu_B \leq 6.5$
    \item 2 functions for the chemical potential dependence of the global fit: linear or parabola
    \item including the coarsest lattice, $N_\tau=8$, or not, in the continuum extrapolation.
\end{itemize}
This amounts to a total of $96=3 \times 2^5$ fits entering the systematic error estimation.

For calculating certain thermodynamic quantities at finite chemical potential, such as the entropy, the temperature 
derivative of the $\lambda_n$ coefficients is also needed. To estimate these derivatives we perform an uncorrelated 
fit of the obtained expansion coefficients with a fourth order polynomial ansatz for the $\lambda^{ij}_2$ and a second order 
polynomial ansatz for the $\lambda^{ij}_4$. For the lower end of our 
temperature range, the fits are constrained to the values predicted by the hadron resonance gas model. In order not 
to dominate the fits, the hadron resonance gas predictions are taken to have error bars of the same size as our lattice 
data. The expansion coefficients and the fits used to estimate their temperature derivative are shown in 
Fig.~\ref{fig:lambdan}. All of the $\lambda_2$ coefficients are approximately constant in the crossover range, 
as is expected from the existence of the approximate scaling variable, discussed in the introduction. 
With the exception of $\lambda^{SS}_2$, the $\lambda_2$ coefficients all go 
to zero within error bars at higher end of our temperature range, as was anticipated 
in the discussion of Fig.~\ref{fig:collapse_SBscheme}.
The $\lambda^{SS}_2$ is still non-zero, as the strangeness 
susceptibility $\chi^S_2$ tends to its Stefan-Boltzmann 
more slowly, due to the 
larger strange quark mass.

\section{Thermodynamics at real $\mu_B$\label{sec:thermo}}
\subsection{Thermodynamics on the $\chi^S_1=0$ line}

\begin{figure*}[t!]
    \begin{center}
        \includegraphics[width=0.47\linewidth]{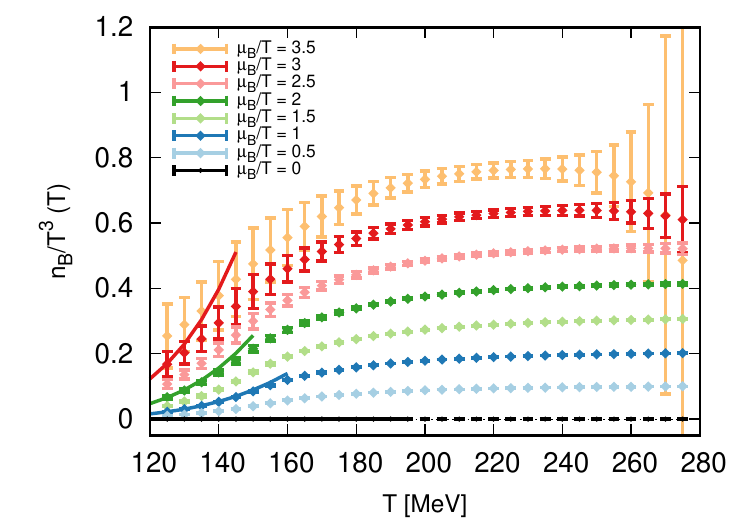}
        \includegraphics[width=0.47\linewidth]{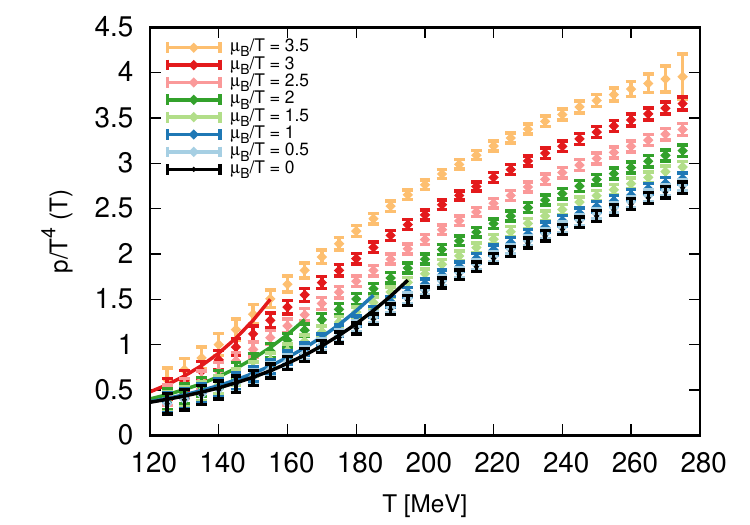} \\
        \includegraphics[width=0.47\linewidth]{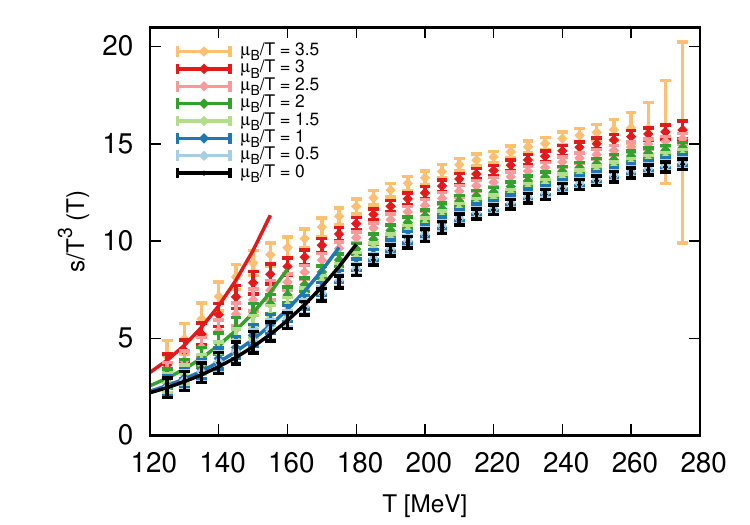}
        \includegraphics[width=0.47\linewidth]{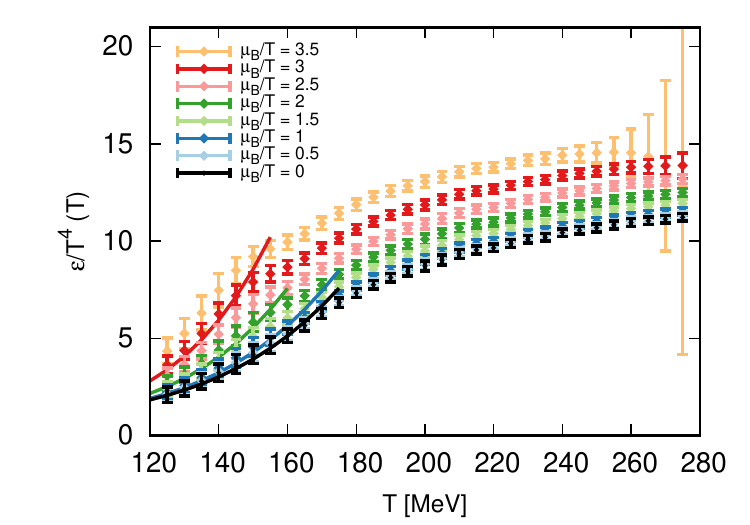} \\
        \includegraphics[width=0.47\linewidth]{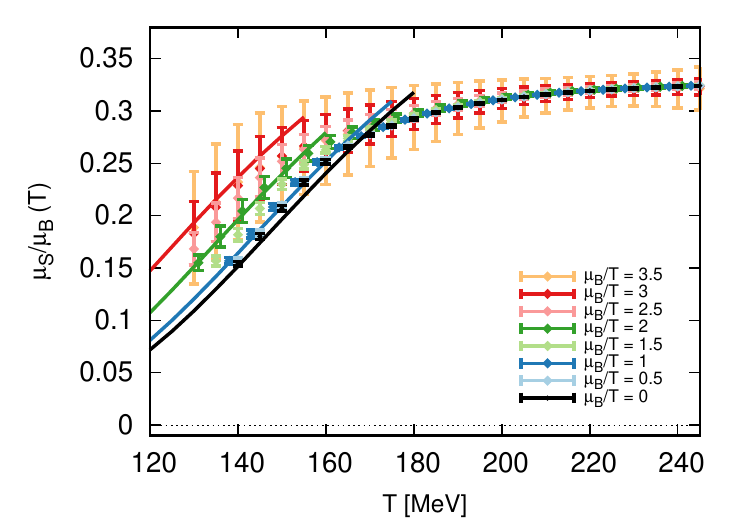}
        \includegraphics[width=0.47\linewidth]{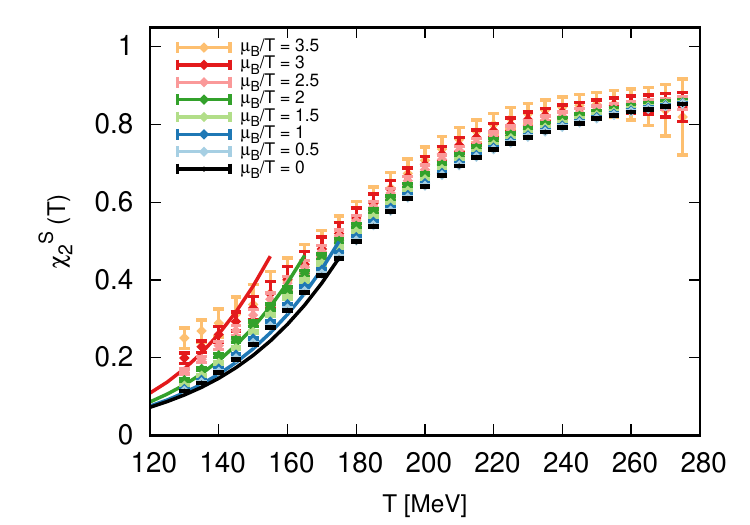}
    \end{center}
    \caption{
    \label{fig:thermo_Sneut}   
    The dimensionless baryon density (top left panel), pressure (top right panel),
    entropy (middle left panel), energy density (middle right panel), strangeness chemical potential-to-baryochemical 
    potential ratio (bottom left panel) and strangeness susceptibility (bottom right panel) as
    functions of temperature at different values of the real chemical potential. The solid lines always show the 
    predictions of the hadron resonance gas model for the corresponding temperature.
    }
\end{figure*}

As we already stated, since $\mu^Q=0$ and $\chi^S_1=0$, on the strangeness neutral line 
we have $c^B_1 = \chi^B_1$. Therefore, the ansatz in Eq.~\eqref{eq:ansatz_BB} gives us the baryon number at 
finite chemical potential directly. In a similar manner, the ratio $\mu_S/\mu_B$ and the strangeness 
susceptibility $\chi^S_2$ are calculated directly from equations~\eqref{eq:ansatz_BS} and~\eqref{eq:ansatz_SS}
respectively. To obtain the pressure, the entropy density and the energy density, we start with the basic quantity 
$c^{B}_1$. For the pressure, one has to calculate the following integral:
\begin{equation}
    \frac{p(T,\hat{\mu}_B)}{T^4} = \frac{p(T,0)}{T^4} + \int_{0}^{\hat{\mu}_B} c^B_{1}(T,\hat{\mu}_B') \dd \hmu_B'\rm{,}
\end{equation}
where
\begin{equation}
    c^B_{1}(T,\hat{\mu}_B) = c^B_{2}(T',0) \frac{\overline{c^B_1}(\hat{\mu}_B)}{\overline{c^B_2}(0)} \rm{,}
\end{equation}
and the pressure at zero chemical potential is taken from Ref.~\cite{Borsanyi:2013bia}. 
The entropy density is defined as 
\begin{equation}
    s = \frac{\partial p}{\partial T} \Bigg|_{\mu_B, \mu_S} \rm{,}
\end{equation}
which can be rewritten in terms of dimensionless quantities as:
\begin{equation}
    \begin{aligned}
    \hat{s} &= 4 \hat{p} + T \frac{\partial  \hat{p} }{\partial T} \Bigg|_{\mu_B} \\
                  &= 4 \hat{p} + T \frac{\partial  \hat{p} }{\partial T} \Bigg|_{\hat{\mu}_B} - \hat{\mu}_B \chi^B_1
                  \rm{,}
    \end{aligned}
\end{equation}
where $\hat{s} \equiv \frac{s}{T^3}$ and we took into account the difference between derivatives at fixed $\mu_B$ versus at fixed $\hat{\mu}_B$. 
By noticing that on the strangeness neutral line 
\begin{equation}
    \begin{aligned}
    \frac{\dd \hat{p}(T,\hmu_B,\hmu_S(T,\hmu_B))}{\dd T} 
    &= \chi^S_1 \frac{\partial \hmu_S}{\partial T} 
       + \frac{\partial \hat{p}}{\partial T} \\
    &= \frac{\partial \hat{p}(T,\hmu_B,\hmu_S(T,\hmu_B))}{\partial T} \rm{,}
    \end{aligned}
\end{equation}
we can write the logarithmic temperature derivative of the pressure as:
\begin{align}
& \quad T  \left. \frac{\partial \hat{p}(T,\hmu_B)}{\partial T} \right|_{\hmu} 
= T   \frac{\partial \hat{p}(T,0)}{\partial T}  \\
& \quad \qquad + \frac12 \int_0^{\hmu_B^2} T \left.\frac{\dd c^B_2(T^\prime,0)}{\dd T^\prime}\right|_{T^\prime = T\left( 1+\lambda_2^{BB} y + \lambda_4^{BB}y^2 \right)} \times \nonumber \\
& \quad \times \left[ 1+\lambda_2^{BB}y +\lambda_4^{BB}y^2 + T \left( \frac{\dd \lambda_2^{BB}}{dT} y+\frac{\dd \lambda_4^{BB}}{dT}y^2 \right) \right] dy \nonumber
\end{align}
where $\frac{\dd c^B_2(T)}{\dd T}$ is calculated at $\mu_B=0$ as discussed previously and presented in Fig.~\ref{fig:c2B_plusderiv}. 

Given the pressure and the entropy, the dimensionless energy density is given by:
\begin{equation}
\hat{\epsilon} = \hat{s} - \hat{p} + \hat{\mu}_B \chi^B_1 \rm{,}
\end{equation}
where $\hat{\epsilon} = \frac{\epsilon}{T^4}$.

The continuum estimates of the dimensionless baryon number, pressure, entropy density, energy density, 
$\mu_S/\mu_B$ ratio and strangeness susceptibility - as computed from the 
expansion coefficients up to order $\lambda^{ij}_4$ - are shown in the various panels of Fig.~\ref{fig:thermo_Sneut}.
Notice that even with the inclusion of the $\lambda^{ij}_4$ coefficients, the statistical errors of our results 
stay well under control in the chemical potential range we study. 
We also compare our results to predictions from the hadron resonance gas model, which at low enough 
temperatures shows an excellent agreement with our lattice data. As the chemical potential increases, 
the crossover temperature decreases, and the agreement between out lattice results and the hadron 
resonance gas gets pushed to lower temperatures, as expected.
Note that, similarly to our previous results for the $\mu_Q=\mu_S=0$ case in Ref.~\cite{Borsanyi:2021sxv}, 
none of the observables display the pathological oscillations of truncated Taylor expansions.

\begin{figure}[t!]
    \begin{center}
        \includegraphics[width=0.98\linewidth]{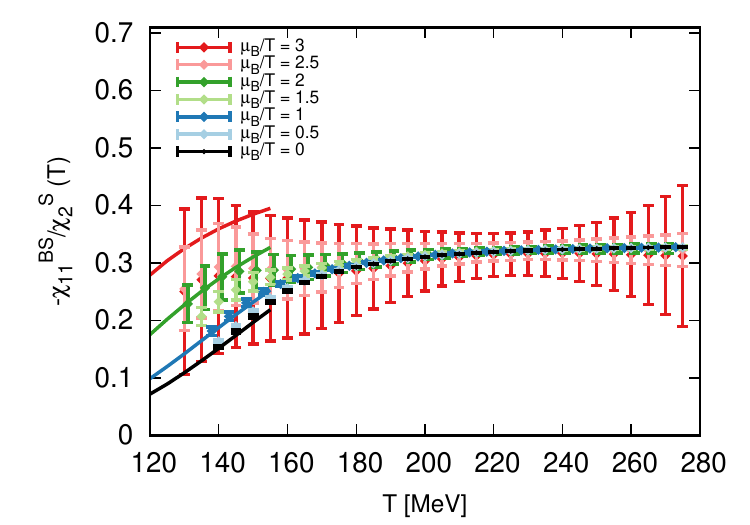}
    \end{center}
    \vspace{-0.45cm}
    \caption{
        \label{fig:BS_SS}
        The ratio $\chi^{BS}_{11}/\chi^S_2$ as a 
        function of the temperature for several values of $\hmu_B$.
    }
\end{figure}

For the extrapolation beyond the strangeness neutral line, the ratio of the baryon-strangeness correlator 
to the strangeness fluctuations $\chi^{BS}_{11}/\chi^{S}_2$ will also 
be needed. In order to obtain this ratio, we simply note that:
\begin{equation}
    \begin{aligned}
    -\frac{\chi^{BS}_{11}}{\chi^S_2} &= \frac{\dd \hmu_S}{ \dd \hmu_B} \\
                                     &= \frac{\dd }{ \dd \hmu_B} \left[ \hmu_B f_{BS}(T_{BS}'(T,\hmu_B)) \right] \\ 
                                     & = \hmu_B \left[ f_{BS}(T_{BS}') + \frac{\partial f_{BS}(T_{BS}'(T,\hmu_B))}{\partial \hmu_B} \right] \rm{,}
    \end{aligned} 
\end{equation}
where we used the shorthand notation $f_{BS}(T) \equiv \rm{M}(T,0) =  -\frac{\chi^{BS}_{11}}{\chi^S_2}(T,\hmu_B=0)$. Our results for 
$\chi^{BS}_{11}/\chi^{S}_2$ are shown in Fig.~\ref{fig:BS_SS}. In addition to being needed for extrapolation to 
finite strangeness, this ratio is also of interest for freeze-out phenomenology~\cite{Bellwied:2019pxh}.

\subsection{Beyond strangeness neutrality}

\begin{figure}[t!]
    \begin{center}
        \includegraphics[width=0.98\linewidth]{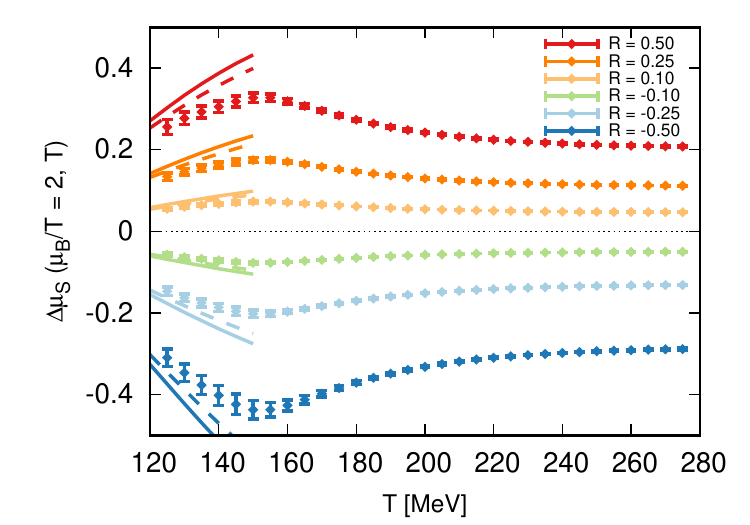}
    \end{center}
    \vspace{-0.45cm}
    \caption{
        \label{fig:Rcomp}
        Shift of the strangeness chemical potential 
        as a function of the temperature at $\hmu_B=2$, at various 
        values of the strangeness-to-baryon ratio $R=\chi^S_1/\chi^B_1$. The solid lines show the exact solution of 
        $0.4 \chi^B_1 = \chi^S_1$ in the hadron resonance (HRG) model, while the dashed lines show the evaluation of the 
        approximation of Eq.~\eqref{eq:DeltaMUS} in the HRG model.
    }
\end{figure}

\begin{figure}[t!]
    \begin{center}
        \includegraphics[width=0.98\linewidth]{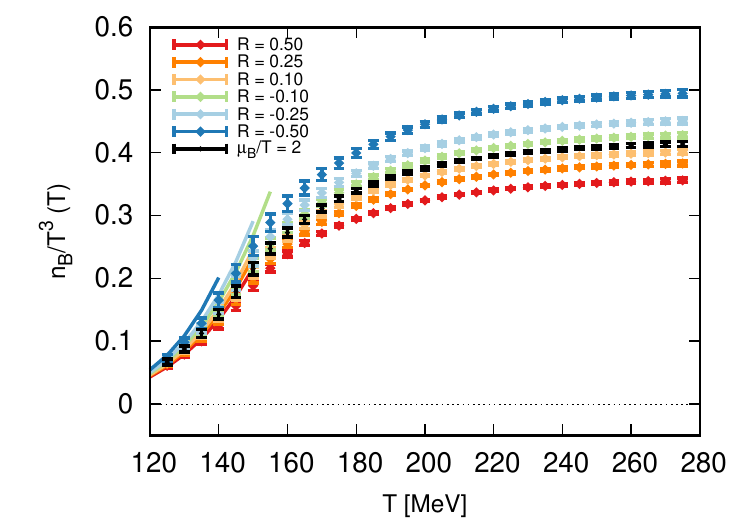}
    \end{center}
    \vspace{-0.45cm}
    \caption{
        \label{fig:nB_beyond}
        The dimensionless baryon density as a function of the temperature at $\hmu_B=2$, for various 
        values of the strangeness-to-baryon  $R=\chi^S_1/\chi^B_1$.
    }
\end{figure}

The quantities calculated so far can also be used to extrapolate the equation of state 
to small values of the strangeness density, slightly off the $\chi^S_1=0$ line.
Let us denote the value of the dimensionless strange quark chemical potential that solves $\chi^S_1=0$ at fixed 
$T$ and $\hmu_B$ as $\hmu_S^*$. Still considering a fixed $\hmu_B$ and $T$, but changing 
$\hmu_S$ slightly from the strangeness neutral choice by a small amount 
\begin{equation}
    \Delta \hmu_S \equiv \hmu_S  - \hmu_S^\star, 
\end{equation}
the dimensionless strangeness and baryon densities become:
\begin{align}
\chi_1^S (\hmu_S) &\approx  \chi_2^S (\hmu_S^\star) \Delta \hmu_S \\
\label{eq:chiB1_beyond}
\chi_1^B (\hmu_S) &\approx \chi_1^B (\hmu_S^\star) + \chi_{11}^{BS} (\hmu_S^\star)  \Delta \hmu_S \rm{,}
\end{align}
where we only kept the linear leading order terms in $\Delta \hmu_S$.
We will express thermodynamic quantities in terms of the strangeness-to-baryon fraction:
\begin{equation}
R = \frac{\chi_1^S}{\chi_1^B} = \frac{ \chi_2^S (\hmu_S^\star) \Delta \hmu_S}{ \chi_1^B (\hmu_S^\star) \Delta \hmu_S + \chi_{11}^{BS} (\hmu_S^\star)} \rm{.}
\end{equation}
Inverting this equation we get:
\begin{equation}
    \label{eq:DeltaMUS}
    \Delta \hmu_S = \frac{ R \hat \chi^B_1 (\hmu_S^\star) }{ \chi_2^S (\hmu_S^\star) - R \chi_{11}^{BS} (\hmu_S^\star)} \rm{.}
\end{equation}
This quantity is shown for $\hmu_B=2$ as a function of temperature for various value of $R$ in Fig.~\ref{fig:Rcomp}.
Substituting Eq.~\eqref{eq:DeltaMUS} into Eq.~\eqref{eq:chiB1_beyond} we obtain - to leading order in $R$:
\begin{equation}
    \frac{\chi^B_1(T,\hmu_B,R)}{\chi^B_1(T,\hmu_B,R=0)} \approx   1 + R \frac{\chi^{BS}_{11}(T,\hmu_B,R=0)}{\chi^S_2(T,\hmu_B,R=0)}  \rm{,}
\end{equation}
where all quantities on the right hand side are along the strangeness neutral line.
We show the results of a leading order extrapolation of the 
dimensionless baryon density as a function of $T$ at $\hmu_B=2$ for several values of $R$ in 
Fig.~\ref{fig:nB_beyond}. 

\begin{figure}[h!]
    \begin{center}
        \includegraphics[width=0.98\linewidth]{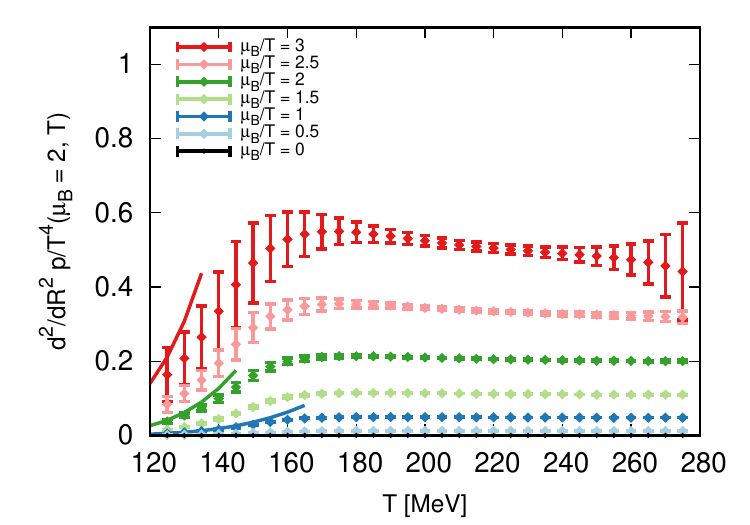}
    \end{center}
    \vspace{-0.45cm}
    \caption{
        \label{fig:pLO_R}
        The leading order Taylor coefficient of the pressure in the strangeness-to-baryon ratio $R$ on the strangeness neutral line 
        as a function of the temperature for several fixed values of $\hmu_B$.
    }
\end{figure}

At the strangeness neutral line the $\mathcal{O}(R)$ correction of the pressure vanishes. The leading order correction gives:
\begin{equation}
    \hat{p}(T,\hmu_B,R) \approx  \hat{p}(T,\hmu_B,R) + \frac{1}{2} \frac{\dd^2 \hat{p}}{\dd R^2}\left( T, \hmu_B \right) R^2\rm{,}
\end{equation}
where 
\begin{equation}
    \frac{\dd^2 \hat{p}}{\dd R^2}\left( T, \hmu_B \right) 
    = \frac{\left( \chi^B_1\left( T, \hmu_B \right) \right)^2}{\chi^S_2\left( T, \hmu_B \right)}\rm{.}
\end{equation}
This leading order coefficient is show in Fig.~\ref{fig:pLO_R}
for several values of $\hmu_B$.

\section{Summary and discussion}

In this work, we generalized the extrapolation procedure we 
introduced in Ref.~\cite{Borsanyi:2021sxv} for the equation of state 
to the case of strangeness neutrality. Using zero and imaginary chemical potential 
simulations we determined the first two coefficients in the scheme $\lambda^{ij}_{2}$ and 
$\lambda^{ij}_{4}$ for the baryon density, the strangeness chemical potential and the 
strangeness fluctuations. Using these coefficients, we extrapolated the equation of state 
up to a baryochemical potential-to-temperature ratio of $\hmu_B=3.5$, considerably improving the range
covered by first principle lattice calculations. Just like for the case 
of $\mu_S=\mu_Q=0$ in Ref.~\cite{Borsanyi:2021sxv}, also for the strangeness neutral case we 
observed that our extrapolation scheme is free of the unphysical oscillations that plague the
fixed order Taylor expansions at higher chemical potentials.

We also introduced a Stefan-Boltzmann correction to improve the rate of convergence of our 
scheme at high temperatures. Due to this correction, the coefficient $\lambda^{BB}_{2}$ goes to 
zero at high temperatures.

We moved beyond strangeness neutrality by calculating the baryon-strangeness correlator to strangeness 
susceptibility ratio $\chi^{BS}_{11}/\chi^S_2$ at finite real $\hmu_B$ on the strangeness neutral line. This allows one
to perform a leading order extrapolation of the baryon number in the ratio $R=\chi^S_1/\chi^B_1$.

Apart from increasing precision through higher statistics,
this work can be extended in the future to to take the small isospin imbalance $\chi^Q_1 = 0.4 \chi^B_1$ of
lead and gold nuclei into account beyond leading order. With significantly more data taking,
a full scan of the $\mu_B-\mu_S$ plane is possible, thus determining the equation of state for higher values of the 
strangeness-to-baryon ratio. It will be valuable to see whether the calculation of even
higher order coefficients (e.g. $\kappa^{ij}_6$) will alter the presented picture or introduce significant improvements
in the range explored by the RHIC Beam Energy Scan program. 

We also mention that, while for the ordinary Taylor expansion of the pressure estimates of the radius 
of convergence exist both on coarse 
lattices~\cite{Giordano:2019slo,Giordano:2019gev,Mondal:2021jxk,Dimopoulos:2021vrk} 
and from phenomenological arguments 
~\cite{Stephanov:2006dn, Mukherjee:2019eou},
the ultimate convergence properties of our scheme are completely unknown at the moment. 
As our ansatz is not a Taylor series for the pressure, 
its domain of convergence is not necessarily a circle 
in the complex plane. Furthermore, the scheme is generalizable also for 
non-polynomial approximations of $T'$, such as rational functions.
The domain of convergence of our scheme is an interesting theoretical question. 
It, however, has no practical consequence at the moment, 
as the $\lambda^{ij}_4$ are already consistent with zero, 
leading to fast apparent convergence at low orders.

\begin{acknowledgments}
    The project was supported by the BMBF Grant No. 05P18PXFCA and 05P21PXFCA.
    This work was also supported by the Hungarian National Research,
    Development and Innovation Office, NKFIH grant KKP126769.
    This material is based upon work supported by the National Science Foundation 
    under grants no. PHY-1654219, PHY-2116686 and OAC-2103680.
    A.P. is supported by the J. Bolyai Research
    Scholarship of the Hungarian Academy of Sciences and by the \'UNKP-21-5 New
    National Excellence Program of the Ministry for Innovation and Technology.
    The authors gratefully acknowledge the Gauss Centre for Supercomputing e.V.
    (www.gauss-centre.eu) for funding this project by providing computing time on the
    GCS Supercomputers HAWK at HLRS, Stuttgart as well as the
    JUWELS/Booster and JURECA/Booster at FZ-Juelich. Part of the computation was performed
    on the QPACE3 funded by the DFG and hosted by JSC.
\end{acknowledgments}

\vspace{0.5cm}

\section*{Appendix: $\lambda^{ij}_2$ coefficients from Taylor coefficients}

For reference, we list here the relationships between our $\lambda_2$ coefficients and ordinary 
Taylor coefficients.
For the expansion coefficient of the baryon density, we get:
\begin{align*}
\lambda_2^{\rm BB} = \frac{1}{6 T f^\prime(T)} \left( c_4^B (0,T) - \frac{\overline{c_4^B}(0)}{\overline{c_2^B}(0)} f (T) \right)
\end{align*}
where $f(T) = \frac{\dd^2 logZ}{\dd \mu_B^2}(\mu_B=0,T)$. 
For the expansion coefficient of the strangeness chemical potential we get
\begin{align*}
\lambda_2^{\rm BS} = \frac{1}{T f^\prime(T)} s_3 (T) = \frac{1}{6 T f^\prime(T)} \frac{\dd^3 \hmu_S}{\dd \hmu_B^3} (T) 
\end{align*}
where $\frac{\hmu_S}{\hmu_B}(\hmu_B,T) = s_1(T) + s_3(T) \hmu_B^2 + s_5(T)\hmu_B^4+\dots$ 
and $f(T) = \lim_{\hmu_B\rightarrow0}\frac{\hmu_S}{\hmu_B}(\mu_B,T) = - \frac{\chi_{11}^{BS}}{\chi_2^S} (0,T)$. 
For the expansion coefficient of the strangeness susceptibility we get
\begin{align*}
\lambda_2^{\rm SS} = \frac{1}{2 \,  T f^\prime(T)} S_{2, \rm sym}^{\rm NLO} (0,T) 
\end{align*}
where $f(T) = \chi_2^S(\mu_B=0,T)$. 

In principle, the $\lambda_4$ coefficients can also be expressed using the Taylor coefficients at $\mu\equiv0$.
For this one needs The Taylor coefficients up to sixth order and the second
temperature derivative of the second order coefficients. For the quantities discussed in this paper we give:

\begin{align}
\lambda_4^{\rm BB}(T) &= \frac{1}{360 T} \frac{1}{{\overline{c}_2^B(0)}^2 {f^\prime (T)}^3} \cdot\nonumber\\
&\left[ 3 \, {\overline{c}_2^B(0)}^2 c_6^B(0,T) {f^\prime (T)}^2  \right. \\ \nonumber
& \left.  - 10 \, \overline{c}_4^B (0) \, {f^\prime (T)}^2 \, \left( \overline{c}_2^B(0) c_4^B (0,T) - \overline{c}_4^B(0) f(T) \right) \right. \\ \nonumber
& \left.  - 5 \, f^{\prime \prime}(T) \left( \overline{c}_2^B(0) c_4^B(0,T) - \overline{c}_4^B(0) f(T) \right)^2 \right] \, \,
\end{align}

\begin{align}
\lambda_4^{\rm BS}(T) &= \frac{s_5(T)}{T f^{\prime}(T)} - \frac{s_3(T)^2 f^{\prime\prime} (T)}{2 T f^\prime(T)^3} \\
&=
\frac{1}{120 T f^{\prime}(T)} \frac{\dd^5 \hmu_S}{\dd \hmu_B^5} (T) - \frac{f^{\prime\prime} (T)}{72 T f^\prime(T)^3} \left(  \frac{\dd^3 \hmu_S}{\dd \hmu_B^3} (T) \right)^2 \, \,\nonumber
\end{align}

\begin{align}
\lambda_4^{\rm SS}(T) &= \frac{1}{24 T f^\prime (T)^3} \Bigl( S_{2, \rm sym}^{\rm NNLO} (0,T) f^\prime(T)^2 \nonumber\\
&- 3 f^{\prime \prime} (T) S_{2, \rm sym}^{\rm NLO} (0,T)^2 \Bigr) \, \,
\end{align}
with
\begin{align}
S_{2, \rm sym}^{\rm NLO} (0,T) &= \chi_{22}^{BS} (0,T) \nonumber\\
&+ 2 s_1(T) \chi_{13}^{BS} (0,T) + s_1(T)^2 \chi_{4}^{S} (0,T) \\
S_{2, \rm sym}^{\rm NNLO} (0,T) &= \chi_{42}^{BS} (0,T) + 4 s_1(T) \chi_{33}^{BS} (0,T) \nonumber\\
&+ 6 s_1(T)^2 \chi_{24}^{BS} (0,T)+ 4 s_1(T)^3 \chi_{15}^{BS} (0,T) \nonumber\\
&+ s_1(T)^4 \chi_{6}^{S} (0,T) + 24 s_3(T) \chi_{13}^{BS} (0,T) \nonumber\\
&+ 24\chi^S_4(0,T) s_1(T)s_3(T)
\end{align}
where we used the expansion coefficients of $\mu_S(\mu_B)$:
\begin{eqnarray}
s_1&=& -\frac{\chi_{11}^{BS}}{\chi_{2}^{S}} \\
s_3&=& -\frac{1}{6\chi_2^{S}} \left[
\chi_{4}^{S}s_1^3+3\chi_{13}^{BS}s_1^2+3\chi_{22}^{BS}s_1+\chi_{31}^{BS}
\right]\\
s_5&=& -\frac{1}{120\chi_2^{S}} \bigl[
+\chi_{6}^{S}s_1^5
+5\chi_{15}^{BS}s_1^4
+10\chi_{24}^{BS}s_1^3 \nonumber\\
&&+60\chi_{4}^{S}s_1^2s_3 +120\chi_{13}^{BS}s_1s_3 +60\chi_{22}^{BS}s_3  \nonumber\\
&&+10\chi_{33}^{BS}s_1^2 +5\chi_{42}^{BS}s_1
+\chi_{51}^{BS}
\bigr]
\end{eqnarray}

\input{eos_.bbl}

\end{document}

%% file: eos_.bbl
\providecommand{\noopsort}[1]{}\providecommand{\singleletter}[1]{#1}%